\PassOptionsToPackage{english}{babel}

\documentclass[%
11pt,
aip,
amsmath,amssymb,
reprint,
floatfix,
]{revtex4-1}

\usepackage[english]{babel}
\usepackage{graphicx}
\usepackage[utf8]{inputenc}
\usepackage{hyperref}
\usepackage{mathtools}
\usepackage[T1]{fontenc} 
\usepackage{xcolor}

\def\sarg{\emph{Sargassum} }
\def\pct{\%}
\def\bx{\ensuremath{\mathbf x}}
\def\bu{\ensuremath{\mathbf u}}
\def\bv{\ensuremath{\mathbf v}}
\def\bw{\ensuremath{\mathbf w}}
\def\be{\ensuremath{\mathbf e}}

\newcommand{\defn}{\coloneqq} 

\begin{document}

\selectlanguage{English}

\title{Physics-informed laboratory estimation of \sarg windage}

\author{M.J.\ Olascoaga} 
\affiliation{Department of Ocean Sciences,  Rosenstiel School of
Marine and Atmospheric Science, University of Miami, Miami, Florida}

\author{F.J.\ Beron-Vera}
\affiliation{Department of Atmospheric Sciences,  Rosenstiel School of
Marine, Atmospheric, and Earth Science, University of Miami, Miami, Florida}

\author{R.T.\ Beyea}
\affiliation{LGL Ecological Research Associates, Inc., Bryan, Texas}

\author{G.\ Bonner}
\affiliation{Department of Atmospheric Sciences,  Rosenstiel School of
Marine, Atmospheric, and Earth Science, University of Miami, Miami, Florida}

\author{M.\ Castellucci}
\affiliation{University of Miami, Coral Gables, Florida}

\author{G.J.\ Goni}
\affiliation{Atlantic Ocean Atmosphere Laboratory,  National Oceanic and Atmospheric Administration, Miami, Florida}

\author{C.\ Guigand}
\affiliation{Department of Ocean Sciences,  Rosenstiel School of
Marine, Atmospheric, and Earth Science, University of Miami, Miami, Florida}

\author{N.F.\ Putman}
\affiliation{LGL Ecological Research Associates, Inc., Bryan, Texas}

\begin{abstract}
A recent Maxey--Riley theory for \sarg raft motion, which models a raft as a network of elastically interacting finite-size, buoyant particles, predicts the carrying flow velocity to be given by the weighted sum of the water and air velocities $(1-\alpha)\mathbf{v} + \alpha \mathbf w$. The theory provides a closed formula for parameter $\alpha$, referred to as \emph{windage}, depending on water-to-particle-density ratio or buoyancy ($\delta$). From a series of laboratory experiments in an air--water stream flume facility under controlled conditions, we estimate $\alpha$ ranging from 0.02 to 0.96\pct. On average, our windage estimates can be up to 9 times smaller than considered in conventional \emph{Sargassum} raft transport modeling, wherein it is customary to add a fraction of $\mathbf w$ to $\mathbf{v}$ chosen in an ad-hoc piecemeal manner. Using the formula provided by the Maxey--Riley theory,  we estimate $\delta$ ranging from 1.00 to 1.49. This is consistent with direct $\delta$ measurements, ranging from 0.9 to 1.25, which provide support for our $\alpha$ estimation.
\end{abstract}


\maketitle

Pelagic \sarg has been known to be abundant throughout the North Atlantic, particularly the western end of the subtropical gyre, known as the Sargasso Sea. Forming rafts, these brown algae serve as an important habitat for ecologically and economically important marine fauna \citep{Bertola-etal-20}. Intriguingly, since 2009 the equatorial Atlantic has emerged as a new region of extreme \sarg abundance \citep{Wang-etal-19,Addico-etal-16}. Pelagic \sarg has since seasonally inundated the coasts of the Caribbean Sea, Gulf of Mexico, South Florida, northern Brazil, and western Africa. Representing a new form of coastal hazard, \sarg inundations have direct impacts on the water quality, nearshore ecosystems, and the local economies \citep{Smetacek-Zingone-13}. Predicting the locations and severity of coastal inundations of pelagic \sarg is a challenging problem.  

The challenge stems in large part from the fact that a \sarg raft's motion is fundamentally unlike Lagrangian (i.e., infinitesimally small, neutrally buoyant) particle motion since it is a finite-size and buoyant object subjected to the action of ocean currents and winds mediated by inertia effects \citep{Beron-21-ND}.

The de-jure fluid mechanics model for describing such effects is provided by the Maxey--Riley equation \citep{Maxey-Riley-83}.  This equation is a Newton-type equation with forces (mainly flow, added mass, and drag) that affect the motion of small, spherical ``inertial'' particles immersed in the flow of a fluid \citep{Cartwright-etal-10}.  For particles floating at the ocean surface a Maxey--Riley equation has been recently proposed by \citet{Beron-etal-19-PoF}. The equation, referred to herein as the \emph{BOM equation}, was verified in the field \citep{Olascoaga-etal-20,Miron-etal-20-GRL} and under controlled laboratory conditions \citep{Miron-etal-20-PoF}.  One aspect, among several others including Earth rotation effects, that makes the BOM equation different than the Maxey--Riley equation is that the carrying flow velocity is given by
\begin{equation}
  \bu(\bx,t) 
  \defn (1-\alpha)\bv(\bx,t) + \alpha \bw(\bx,t).
  \label{eq:u}
\end{equation}
Here, $\bv(\bx,t)$ and $\bw(\bx,t)$ ($\bx$ and $t$ denote horizontal position and time, respectively) are water and air velocities near the water surface, respectively, and $\alpha \in [0, 1)$ is a coefficient that depends in closed form on the water-to-particle-density ratio, which we denote by $\delta$ and call \emph{buoyancy} (the reserve volume is $1-\delta^{-1}$) (cf.\@~Appendix \ref{app:alpha}). We call $\alpha$ itself the \emph{windage}; it is a fractional measure of the overall contribution of wind to the carrying flow velocity in the BOM equation.

Buoyancy-dependent windage $\alpha(\delta)$ has been identified as the most important factor controlling isolated object drift at the ocean surface \citep{Olascoaga-etal-20,Miron-etal-20-GRL} and can be expected to have a similarly influential role in \sarg raft drift. In this note we seek to frame \sarg windage in the laboratory under controlled conditions by building on the BOM model.  

The BOM equation, however, cannot be expected to describe \sarg raft motion.  A Maxey--Riley model for the drift of \sarg rafts was proposed by \citet{Beron-Miron-20} based on the BOM equation by envisioning them as networks of elastically interacting inertial particles. The gas-filled bladders that keep a raft afloat represent the inertial particles in the proposed model, herein referred to as the \emph{eBOM model}, and the flexible stems that connect them are substituted by massless springs.  While the eBOM model still requires quantitative testing against observations, it has been successful in qualitatively explaining transport of \sarg and coastal inundation in the Caribbean Sea \citep{Andrade-etal-22}.

The eBOM model is a second-order system of ordinary differential equations coupled by the linear-elastic spring forces acting between adjacent particles of the network. It is assumed that each particle and spring is identical.  However, in the nonrotating case with constant $\bv$ and $\bw$, and hence $\bu$ in Eq.\@~\eqref{eq:u}, the motion of the $i$th inertial particle of an elastically interacting network, i.e., a \sarg raft, obeys
\begin{equation}
  \ddot\bx_i = \tau^{-1}(\bu - \dot\bx_i),
  \label{eq:eBOM}
\end{equation}
where $\tau$, proportional to the particle radius squared, is the Stokes time, measuring the inertial response time of the particle to the two-component water--air medium (cf.\ Appendix \ref{app:alpha}). The assumption of constant $\bu$ removes all of its time derivatives from the eBOM model, but the spring forces between particles still remain. However, Eq.\@~\eqref{eq:eBOM} is possible since the elastic forces acting on the $i$th particle will vanish for all $t$ so long as it is initially separated from its neighbors by a distance equal to the natural length of the connecting springs.  This suggests that, in this special case and with the appropriate initial conditions, \sarg raft windage can be estimated from measurements of raft velocity $\bv_i \defn \dot\bx_i$, satisfying
\begin{equation}
  \bv_i(t) = (\bv_i(0) -\bu) \exp(-t/\tau) + \bu. 
  \label{eq:vi}
\end{equation}
Indeed, if $\bv_i(0)=\bu$ then $\bv_i(t) = \bu$ for all $t$, allowing one to estimate $\alpha$ from measurements of $\bv_i$ given $\bv$ and $\bw$ using \eqref{eq:u}. (We take the opportunity to correct a mistake incurred in \citet{Miron-etal-20-PoF}: Eq.\@~(2), there, must be replaced by Eq.\@~\eqref{eq:vi}, here, with $i$ replaced by p, and ``was negligible'' in the 2nd para.\ of the left col.\ in p.\ 3 must read ``can be expected to be close to the carrying flow speed.'')  Below we report \sarg windage estimates obtained from a series of laboratory experiments where the \sarg raft velocity was measured as rafts drifted in a flume with constant water velocity and air streams of varied intensities.

\begin{figure}[h]
  \centering%
  \includegraphics[width=\columnwidth]{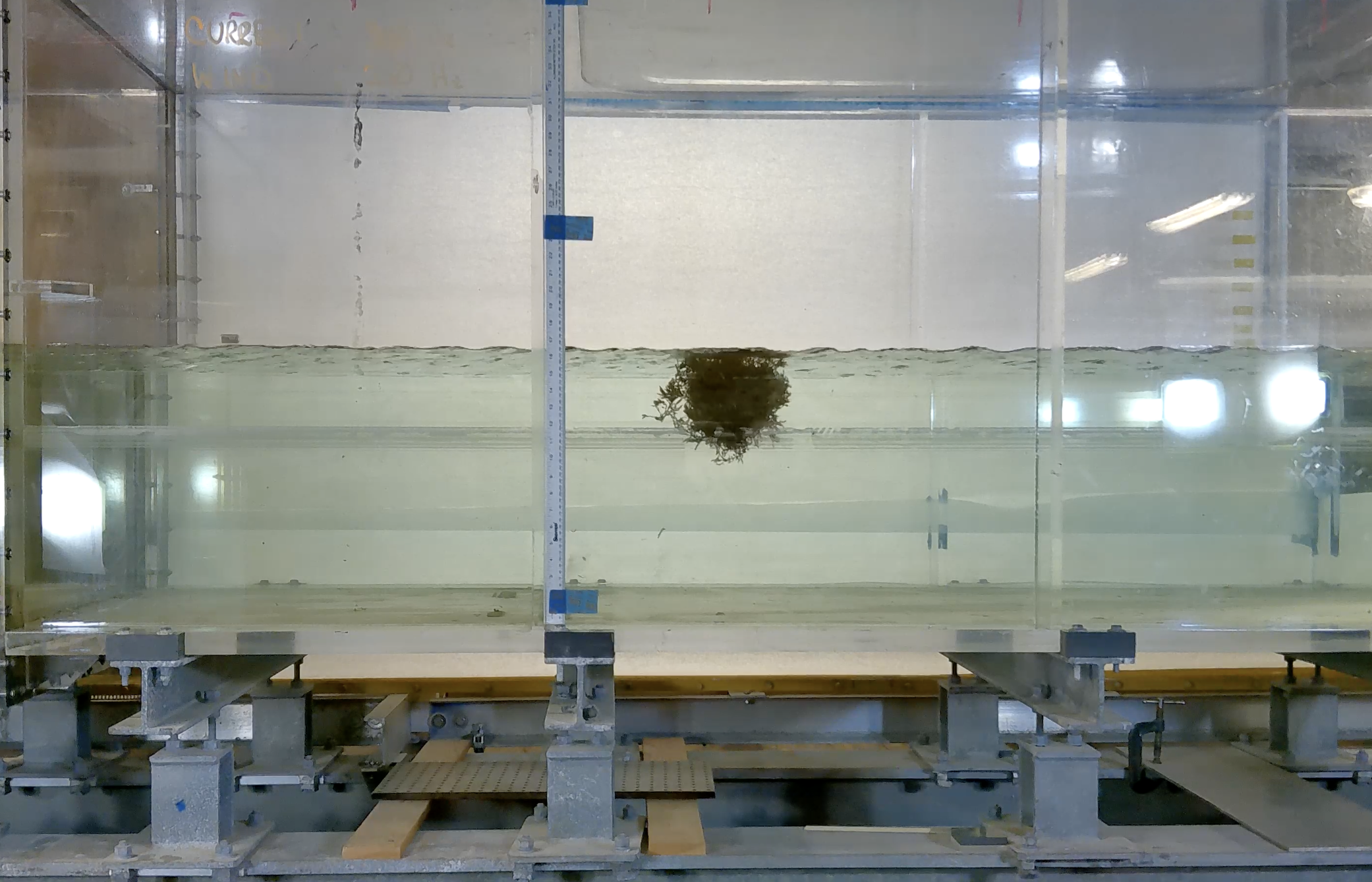}%
  \caption{Side view of the ASIST air--water stream flume where the \sarg windage experiments were carried out. A \sarg cluster, representative of those employed in the experiments, is shown floating at the water level.  The clusters used were composed of clumps collected from a raft found in the Florida Current off the southeastern coast of the Florida Peninsula.}
  \label{fig:flume}%
\end{figure}

The laboratory experiments were carried out in the Air--Sea Interaction Salt--water Tank (ASIST) of the Alfred G. Glassell, Jr.\ SUrge STructure Atmosphere INteraction (SUSTAIN) facility of the University of Miami's Rosenstiel School of Marine, Atmospheric, and Earth Science (\url{https://sustain.rsmas.miami.edu/}) (Fig.\ \ref{fig:flume}). ASIST allows to control the water stream with a pump and the air stream with a fan. The acrylic ASIST flume is 15-m long, with a cross section of 1 m $\times$ 1 m.  In our experiments, the flume was filled with seawater, of density 1.020 gr\,cm$^{-3}$, up to reaching a (mean) depth of 0.43 m.  The flume bottom included a gentle slope at the head to damp wave reflection.

\begin{figure}[h]
  \centering%
  \includegraphics[width=\columnwidth]{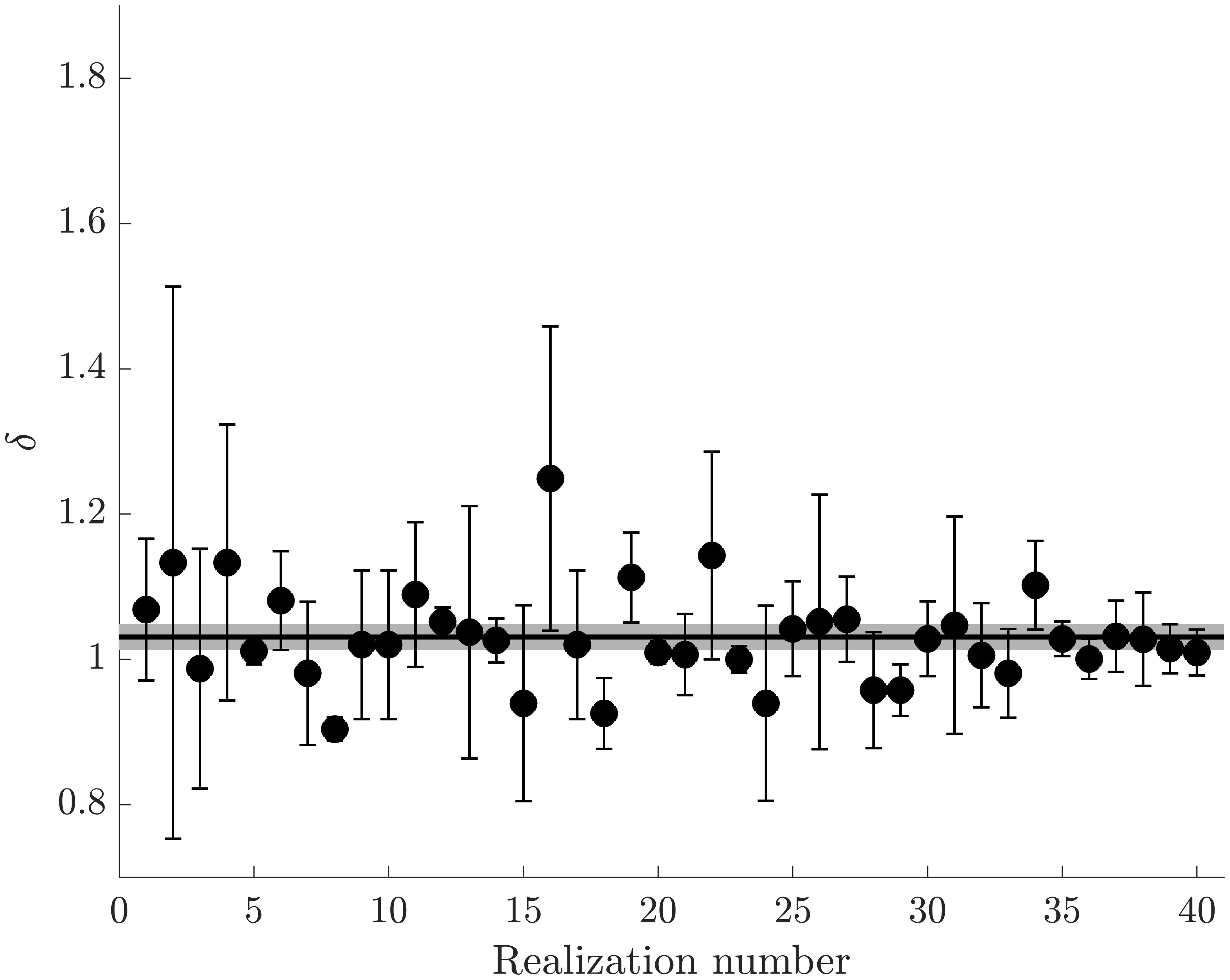}%
  \caption{For each cluster of \sarg formed by a group of clumps chose at random, a dot represents a buoyancy estimate obtained by measuring the mass and volume of the cluster. The accompanying error bar is the result of propagating the mass and volume measurement errors. The thick horizontal line is the weighted average of the individual $\delta$ estimates, where the surrounding shade represents a one-standard-deviation uncertainty.}
  \label{fig:delta_hist}%
\end{figure}

Clumps, i.e., individual plants, of \sarg were collected in the Florida Current off the southeastern coast of the Florida Peninsula from a small raft, with a horizontal areal coverage of about 0.5 m $\times$ 0.5 m and thickness of approximately 10 cm. Only a small portion of the raft was above water at the time of collection.  The collected \sarg clumps belonged to various species, mainly \emph{Fluitans} and \emph{Natans I} and \emph{VII}.  We estimated the density of several clusters formed by a group of clumps selected at random.  Each cluster was weighted and its volume computed from the displaced volume using graduated cylinders of two different sizes. The dots in Fig.\@~\ref{fig:delta_hist} are the $\delta$ values estimated for the clusters, ranging approximately from 0.91 to 1.25.  The accompanying error bars are obtained by propagating the mass and volume measurement errors while ignoring those associated with the seawater density, using the standard formula which treats them as independent and random.\citep{Ripa-CM-02} The mass measurement error was 0.1 gr, while those of the volume were 1 and 5 ml when made using the smaller and larger cylinder, respectively. Performing a least squares fit of the data to a constant, we obtained $\delta = 1.03 \pm 0.02$. This followed by minimizing the weighted sum of the residuals squared with the weights given by the inverse of the data errors, representing our confidence on the data, and then assuming that the data had errors which were independent and random.\citep{Ripa-CM-02}  The result is a weighted mean with an uncertainty given by the weighted standard deviation of the data divided by the square root of the number of data minus one, the fit's degrees of freedom.  We apply a similar procedure below, loosely referring to it as an error propagation procedure.  The result that $\delta \gtrapprox 1$ is consistent with the visual observation that most clumps were at the ocean water level at the time of collection.

Five fan frequencies were considered in the ASIST flume experiments, 15, 20, 25, 30, and 35 Hz.  These correspond \citep{Novelli-etal-17} to air stream speeds (extrapolated to the mean water level from measurements roughly 0.3 m above using bulk formulas \citep{Smith-88, Hsu-etal-94}) of approximately 410, 580, 760, 950 and 1150 cm\,s$^{-1}$, respectively.  The resulting air stream speeds produced negligible to small ripples at the air--water interface, except for the highest two, which led to higher amplitude waves possibly inducing drift. The water pump frequency was set to 30 Hz.  Due to friction with the walls and bottom, and the stress applied by the air stream at the surface, the resulting water stream is necessarily sheared, mainly in the vertical.  Intensification of the water stream toward the surface depends on the intensity of the air stream.  This was observed from particle image velocimetry (PIV) by \citet{Novelli-etal-17} in an experimental setup that we followed as closely as possible.  An effective water stream speed, depending on the air stream speed in each experiment conducted, was obtained by vertically averaging over the \sarg cluster submerged depth the velocity profile inferred from PIV. \citet{Novelli-etal-17} note that PIV measurements are not reliable within the top 5 cm of the water column; extrapolated values by \citet{Novelli-etal-17} in that layer were not taken into account.  We have not considered all PIV profiles reported by \citet{Novelli-etal-17}, as lack of correspondence at some depth levels with the applied air stream was noted.  Thus we only considered the velocity profiles resulting at the lowest and highest air stream speeds, and linearly interpolated values at each depth for intermediate speeds.  The resulting vertically averaged water stream speeds, approximately 11.6, 11.7, 12.3, 12.9, 13.5, and 14.0 cm\,s$^{-1}$, increase with increasing air stream speed as can be expected.  In all cases the air and water stream speeds used were representative of ocean current speeds and wind intensities typically observed in the open ocean under nonrough sea conditions. 

\begin{figure}[h]
  \centering%
  \includegraphics[width=\columnwidth]{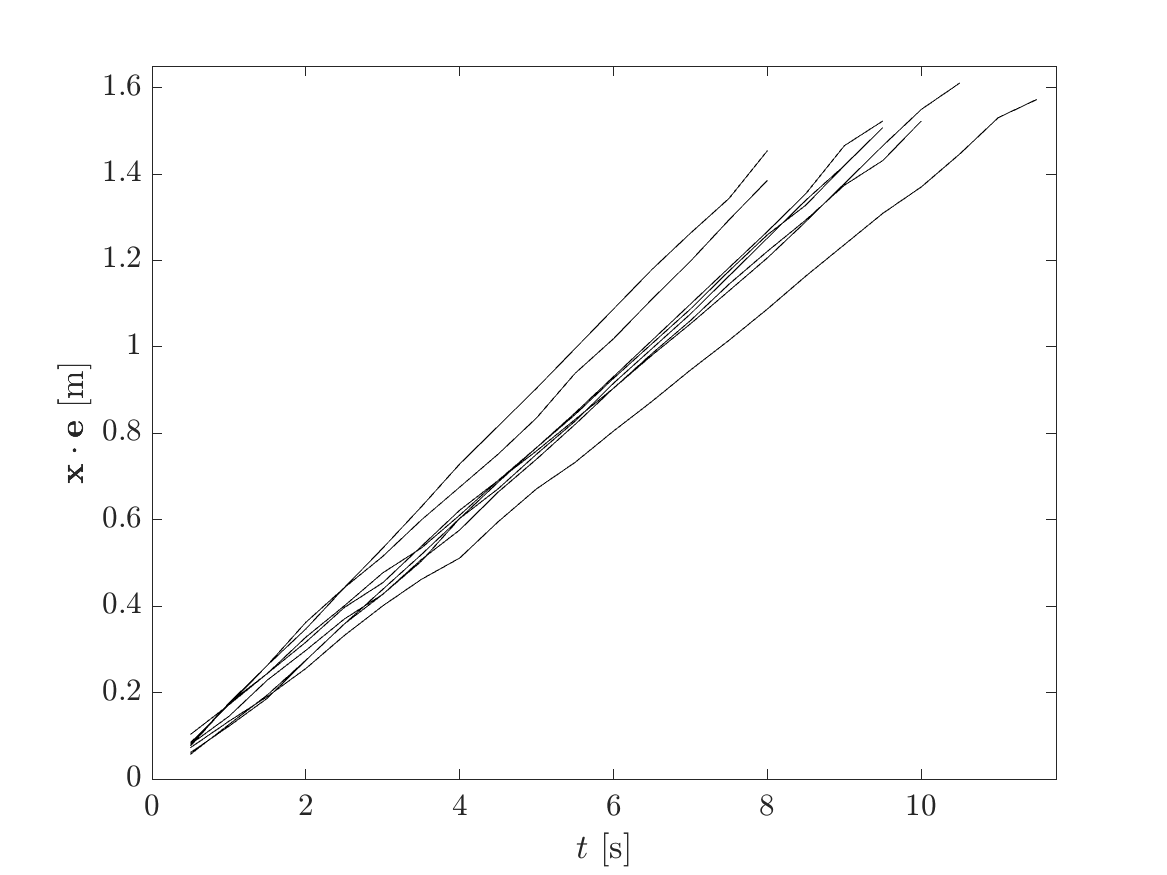}%
  \caption{Distance traveled along the ASIST flume as a function of time by each \sarg cluster in an experiment at fan frequency 25 Hz (7.6-m\,s$^{-1}$ air stream intensity) inferred using CSR-DCF motion tracking indicating approximately constant drift speed.}
  \label{fig:distance}%
\end{figure}

A single experiment consisted in placing a hand-size cluster of \sarg on the water surface and allowing it to drift freely down the flume.  Ten experimental realizations per air stream intensity were conducted.  A few realizations were discarded in which a cluster either floated too closely to the edges of the flume or sank too far below the surface. To estimate a cluster speed, the cluster was first allowed to travel a set distance to ensure that $\bv_i(0) \approx \bu$ by Eq.\@~\eqref{eq:vi}. Then, we measured the time it took the cluster to travel a fixed length of 2.3 m using two independent chronometer measurements.  We also video recorded the experiments and tracked the motion of the clusters using CSR-DCF (Discriminative Correlation Filter with Channel and Spatial Reliability) \citep{Lukezic-etal-18}, which improves the reliability of nonrectangular object tracking.  This verified the chronometer-based estimates of speed, which were largely constant during each realization.  In Fig.\@~\ref{fig:distance} we show distance traveled by each \sarg cluster in the along-flume direction ($\be$) in an experiment at fan frequency 25 Hz ($\bw\cdot\be  \approx 760$ cm\,s$^{-1}$); similar results were seen in experiments at other frequencies.  This observation that $\bv_i(t)$ is approximately constant allowed us to proceed to estimate \sarg windage as proposed.  Variations of speed estimates across experiment realizations were observed. Despite the care taken to ensure the drifting \sarg avoided the walls, lateral flume boundary effects probably still existed. In addition, buoyancy variations across experiments were observed. Indeed, a tendency of the \sarg clumps to lose bladders was noted each time a cluster was removed from the water at the end of an experiment realization to be reused in the subsequent one.

\begin{figure}[h]
  \centering%
  \includegraphics[width=\columnwidth]{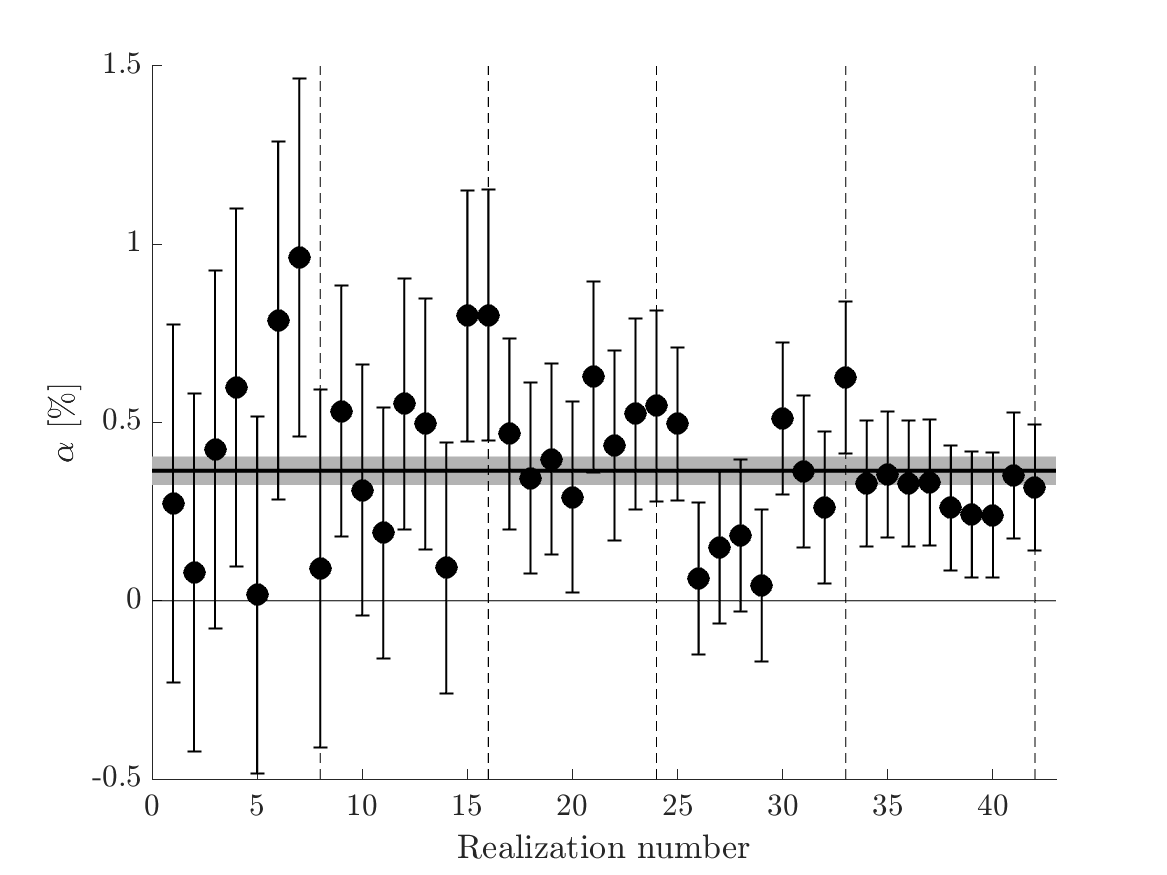}%
  \caption{For each experiment realization a dot represents an estimate of the the windage ($\alpha$) parameter computed with Eq.\@~\eqref{eq:alpha_v_w} and whose error bar accounts for uncertainty in the water and wind speeds via propagation. The thick horizontal line is the weighted average of the individual $\alpha$ estimates with the corresponding uncertainty shaded around it. The dashed lines bound experiment realizations carried out at different wind speeds, increasing to the right.}
  \label{fig:alpha}%
\end{figure}

Each dot in Fig.\@~\ref{fig:alpha} is 
\begin{equation} 
    \alpha = \frac{(\bv_i-\bv)\cdot\be}{(\bw - \bv)\cdot\be}
    \label{eq:alpha_v_w}
\end{equation}
as estimated in a given experiment realization. The broken lines separate experiment realizations conducted at different wind speeds, increasing to the right. The error bars accompanying the estimates, ranging approximately from 0.02 to 0.96\pct, represent their uncertainties computed by propagating the error associated with the water and wind speeds.  These are reported \citep{Novelli-etal-17} to be of 2 cm\,s$^{-1}$ and 1\pct\, per fan frequency, respectively.  The dispersion is larger for slower winds and smaller for stronger winds.  Upon performing a weighted average over all individual $\alpha$ estimates we obtain $\alpha = (0.36 \pm 0.04)\pct$, where the uncertainty associated with the overall estimate is the result of propagating that associated with each individual estimate.  The result is depicted in Fig.\ \ref{fig:alpha} by the thick horizontal line and accompanying shade.

The reported windage estimate are smaller than the 1--3\pct\ considered in conventional \sarg raft transport modeling \citep{Johns-etal-20, Putman-etal-20, Jouanno-etal-21a}.  However, an application of the BOM model involving a \emph{Sargassum}-like plastic raft tracked by satellite along with satellite-altimetry-derived ocean currents and reanalyzed near-surface winds \citep{Olascoaga-etal-20} appears to favor a lower windage.  Indeed, the trajectory of the artificial \sarg raft, about 2-cm thick and spanning an area of approximately 250 cm $\times$ 50 cm, was best described by the BOM model when its parameters were computed using $\delta = 1.25$, corresponding to $\alpha \approx 0.5\pct$. On the other hand, an analysis involving satellite-tracked trajectories of undrogued surface drifting buoys \cite{Johnson-etal-20} favors the use of a lower windage than in conventional \sarg modeling to improve \sarg connectivity throughout the Tropical Atlantic and achieve a wider spread of \sarg in the eastern Tropical Atlantic than is currently observed in satellite imagery, adhering to earlier hypotheses \cite{Franks-etal-16}.

\begin{figure}[h]
  \centering%
  \includegraphics[width=\columnwidth]{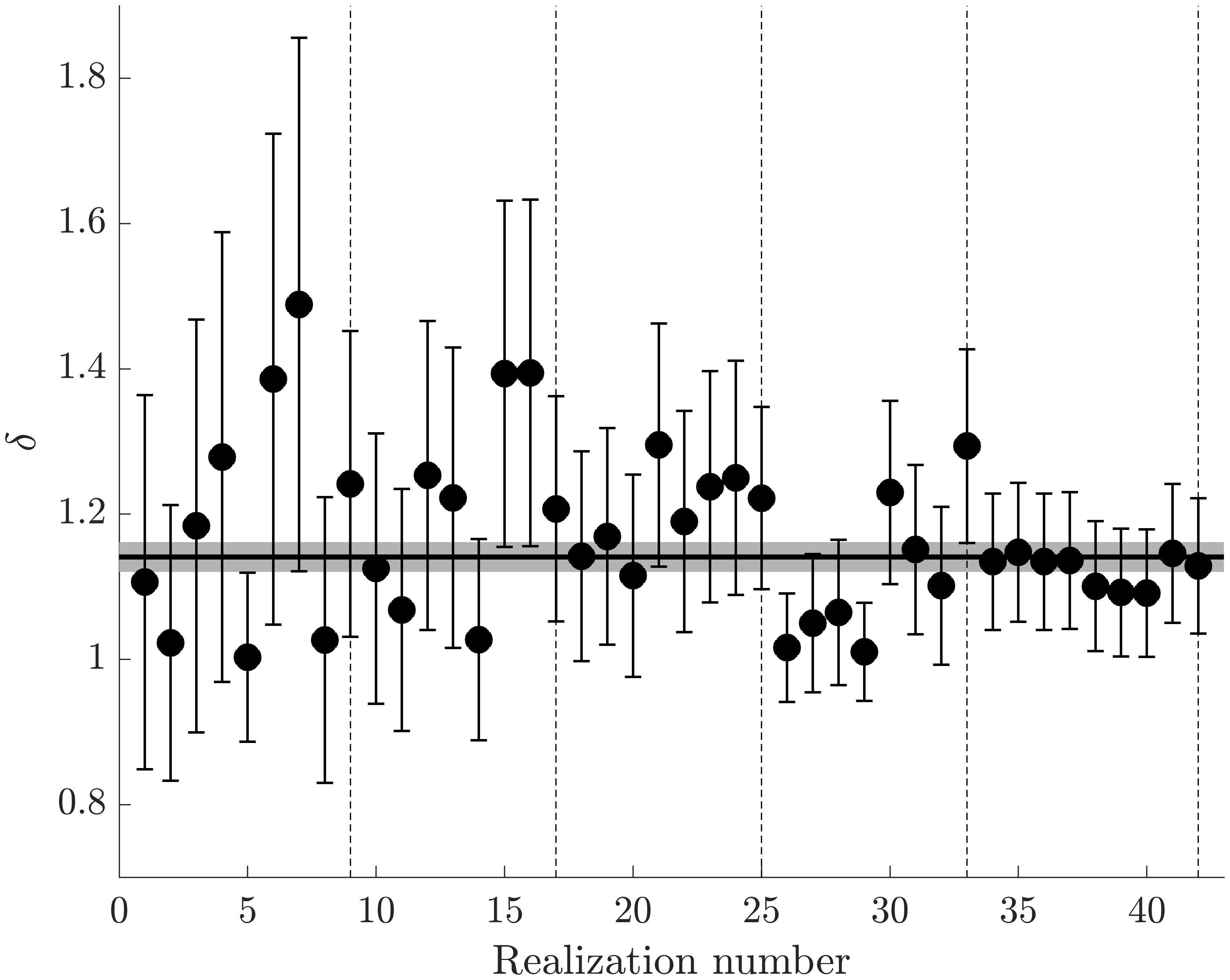}%
  \caption{For each experiment realization a dot represents a buoyancy estimate computed from corresponding windage ($\alpha$) estimate as in Appendix \ref{app:alpha} with an error bar obtained via propagation of the error associated with $\alpha$. The thick horizontal line is the weighted average of the individual $\delta$ estimates with the corresponding uncertainty shaded around it. As in Fig.\@~\ref{fig:alpha}, the dashed lines delimit experiment realizations carried out at different wind speeds, which increase to the right.}
  \label{fig:delta}%
\end{figure}

The reliability of the windage estimates can be better assessed by comparing the corresponding buoyancies with those directly inferred from Fig.\@~\ref{fig:delta}.  As noted above, both the BOM and eBOM model relate windage ($\alpha$) with buoyancy ($\delta$) in closed form.  An approximate inverse relationship, $\delta(\alpha)$, appropriate for use in the near-neutrally buoyant limit, is given in Appendix \ref{app:alpha}.  Figure \ref{fig:delta} shows the resulting $\delta$ estimates depending on the experiment realization, ranging from 1.00 to 1.49.  The overall buoyancy estimate $\delta = 1.14\pm 0.02$ lies within the range of the directly measured $\delta$, albeit toward its larger end. Differences may be attributed, in part, to the direct buoyancy measurements being made using different individual \sarg clusters than those used during the travel time measurements (although both experiments used clusters from the same group of clumps.) This corroboration increases the confidence in the reported windage estimates and, as a consequence, the eBOM modeling framework employed to obtain them.

Our study is the first to directly measure windage of Sargassum under controlled conditions. Given the wide range of windage values used in modeling \sarg transport, and the potential uncertainty this introduces in predictions,\citep{Putman-etal-18, Wang-etal-19, Berline-etal-20, Johns-etal-20, Johnson-etal-20, Jouanno-etal-21a, Jouanno-etal-21b, Alleyne-etal-23, Podlejski-etal-23, Lara-etal-24} direct measurements of wind effects on \sarg movement are of considerable importance for predicting its movement and forecasting beaching events.\citep{Putman-etal-23} We expect to use the estimated windage and corresponding buoyancy values in testing the eBOM model in the field against real \sarg raft trajectories.  We have recently deployed satellite trackers in several \sarg rafts found in the Florida Current off the southern Florida Peninsula to carry out such a test.  The deployments were done in pairs so the eBOM model can be tested, approximately at least.  We will report on results of the test elsewhere. 

\begin{acknowledgments}
We thank Sanchit Mehta, Katherine Simi, and Peisen Tan for helping with the laboratory experiments. This work was supported by NSF grant OCE2148499.
\end{acknowledgments}

\section*{Author declarations}

\subsection*{Conflict of interest}

The authors have no conflicts to disclose.

\subsection*{Author contributions}

Maria Olascoaga: Conceptualization (equal); Data curation (equal);
Formal analysis (equal); Writing - original draft (equal); Writing
- review \& editing (equal). Francisco Beron-Vera: Conceptualization
(equal); Formal analysis (equal); Writing - original draft (equal);
Writing - review \& editing (equal). Taylor Beyea: Data curation
(equal); Writing - review \& editing (equal). Gage Bonner:
Conceptualization (equal); Formal analysis (equal); Writing - review
\& editing (equal). Michael Castellucci: Data curation (equal);
Visualization (equal). Gustavo Goni: Resources (equal); Writing -
review \& editing (equal). Cedric Guigand:  Resources (equal);
Nathan Putman: Data curation (equal); Writing - review \& editing
(equal).

\section*{Data availability}

Video records of the experiments are available from \url{https://github.com/SargassumLab/SargassumWindageVideos}.

\appendix

\section{Buoyancy dependent windage and Stokes time\label{app:alpha}}

According to the BOM equation, the windage ($\alpha$) for spherical particle floating at the ocean--atmosphere interface varies with buoyancy ($\delta$) as
\begin{equation}
  \alpha(\delta) = \frac{\gamma\Psi(\delta)}{1 + (\gamma - 1)\Psi(\delta)}.
  \label{eq:alpha}
\end{equation}
Here, $\gamma$ is the air-to-water viscosity ratio, 
\begin{equation}
  \Psi(\delta) \defn \pi^{-1}\operatorname{acos}\Phi(\delta) - \pi^{-1}\Phi(\delta)\smash{\sqrt{1 - \Phi(\delta)^2}},
  \label{eq:Psi}
\end{equation}
giving the fraction of emerged particle's projected (in the flow direction) area,
\begin{equation}
  \Phi(\delta) \defn
  \frac{1}{2}(\varphi(\delta)^{-1} + \varphi(\delta))
  +
  \frac{\mathrm{i}\sqrt{3}}{2}
  \left(\varphi(\delta) - \varphi(\delta)^{-1}\right),
\end{equation}
with the fraction of emerged particle piece's height given by $1 - \Phi(\delta)$, where
\begin{equation}
  \varphi(\delta) \defn \sqrt[3]{\mathrm{i}\sqrt{1 - (2\delta^{-1}
  - 1)^2} + 2\delta^{-1} - 1}.
\end{equation}
Note the slight change of notation in Eq.\@~\eqref{eq:Psi} with respect to earlier references\citep{Miron-etal-20-GRL, Miron-etal-20-PoF, Beron-Miron-20, Beron-etal-19-PoF, Olascoaga-etal-20, Beron-21-ND}.

As pointed out in \citet{Olascoaga-etal-20}, the above formulae is not valid for $\delta \ge 1$ very large, which is of no consequence in general and in particular in the near-neutrally buoyant case, $\delta \approx 1$, which of interest here. We can construct a series reversion for $\delta(\Psi)$, which is accurate to $0.05\%$ or less for $1 \leq \delta \lessapprox 1.5$,
\begin{align} 
    \delta(\Psi) &\sim 1 \notag \\
    &+ 1.481344653620955\cdot \Psi^{4/3} \notag \\
    &- 0.2775826237806382\cdot \Psi^2 \notag \\
    &+ 2.114966253909641\cdot \Psi^{8/3} \notag \\ 
    &- 0.8594639386944948\cdot \Psi^{10/3} \notag \\ 
    &+ 3.071255172923317\cdot \Psi^4 \notag \\ 
    &- 1.906635056270554\cdot \Psi^{14/3} \notag \\ 
    &+ 4.589911692676772\cdot  \Psi^{16/3} \notag \\
    &- 3.708388371098538\cdot  \Psi^6.
    \label{eq:delta-Psi-series}
\end{align}
Solving Eq.\@~\eqref{eq:alpha} for $\Psi(\alpha)$ and substituting the result into Eq.\@~\eqref{eq:delta-Psi-series} provides an approximate formula for $\delta(\alpha)$.

Finally, the Stokes time involved in Eqs.\@~\eqref{eq:eBOM}--\eqref{eq:vi} is defined by
\begin{equation}
    \tau := \frac{a^2\rho}{\mu}\cdot \frac{1 - \frac{1}{6}\Phi}{\big(1 + (\gamma - 1)\Psi\big)\delta^4},
    \label{eq:tau}
\end{equation}
where $a$ is the particle radius, $\rho$ is the water density, and $\mu$ stands for viscosity.

(We note that in three references \citep{Miron-etal-20-GRL, Miron-etal-20-PoF, Beron-Miron-20} it has been typed $1-\gamma$ in Eqs.\@~\eqref{eq:alpha} and \eqref{eq:tau}, instead of $\gamma - 1$, as it must be.)

%

\end{document}